\documentclass{article}
\usepackage{spconf,amsmath,epsfig}

\usepackage[switch]{lineno}
\usepackage{amsmath,amsfonts,graphicx,float}
\usepackage{array}
\usepackage{cleveref}
\usepackage{booktabs}
\usepackage{ragged2e}
\usepackage{multicol}
\usepackage{multirow,bigstrut}
\usepackage{color}
\usepackage{xspace}
\usepackage{graphicx}
\usepackage[justification=centering]{caption}
\usepackage{subcaption}
\usepackage{lipsum}
\usepackage{cite}
\usepackage{lipsum}
\usepackage{blindtext}
\usepackage{multicol}
\usepackage{hhline}
\usepackage{bm}

\begin{document}\sloppy

% Example definitions.
% --------------------
\def\x{{\mathbf x}}
\def\L{{\cal L}}

% Title.
% ------
\title{Modeling Continuous Video QoE Evolution: A State Space Approach}
%
% Single address.
% ---------------

\name{Nagabhushan Eswara$^1$, Hemanth P. Sethuram$^2$, Soumen Chakraborty$^2$,}\secondlinename{Kiran Kuchi$^1$, Abhinav Kumar$^1$, Sumohana S. Channappayya$^1$}

\address{$^1$\normalsize Department of Electrical Engineering, Indian Institute of Technology Hyderabad, India \\
$^2$\normalsize Intel Corporation, India \\
$^1$\normalsize\{ee13p1004, kkuchi, abhinavkumar, sumohana\}@iith.ac.in, \large$^2$\normalsize\{hemanth.p.sethuram,soumen.chakraborty\}@intel.com
\vspace{-0.18in}
}

\maketitle

\begin{abstract}

A rapid increase in the video traffic together with an increasing demand for higher quality videos
% such as ultra high definition 
has put a significant load on content delivery networks in the recent years. Due to the relatively limited delivery infrastructure, the video users in HTTP streaming often encounter dynamically varying quality over time due to rate adaptation, while the delays in video packet arrivals result in rebuffering events. The user quality-of-experience (QoE) degrades and varies with time because of these factors. Thus, it is imperative to monitor the QoE continuously in order to minimize these degradations and deliver an optimized QoE to the users. Towards this end, we propose a nonlinear state space model for efficiently and effectively predicting the user QoE on a continuous time basis. The QoE prediction using the proposed approach relies on a state space that is defined by a set of carefully chosen time varying QoE determining features.
% corresponding to current and previously rendered video time instants. 
An evaluation of the proposed approach conducted on two publicly available continuous QoE databases shows a superior QoE prediction performance over the state-of-the-art QoE modeling approaches. %Thus, demonstrating the effectiveness of the proposed model for QoE prediction.
The evaluation results also demonstrate the efficacy of the selected features and the model order employed for predicting the QoE. Finally, we show that the proposed model is completely state controllable and observable, so that the potential of state space modeling approaches can be exploited for further improving QoE prediction.
%Finally, we show that a simple averaging of continuous time QoE scores can quantify the overall QoE of the users sufficiently well at the end of video viewing.

\end{abstract}


\vspace{-0.1in}

\section{Introduction}
\label{sec:intro}

\vspace{-0.05in}

Streaming videos on demand over Hyper Text Transfer Protocol (HTTP) has grown significantly in the recent years. According to Cisco's VNI \cite{cisco}, videos accounted for 60\% of the total mobile data traffic in 2016. It is projected that more than three-fourth of the world's mobile data traffic will be constituted by videos by 2021. Such a massive growth in the video traffic is putting a huge stress on the video delivery infrastructure. 
%Therefore, it is important for the networks to perform a careful and optimal utilization of the available resources for video streaming while maintaining an acceptable level of Quality-of-Experience (QoE) of the video users. 

In case of video streaming, a large volume of network traffic can cause impairments such as congestion and packet drops which in turn can result in significant delays in the packet arrival at the end user causing the playback to stall. Such events are referred to as rebuffering events \cite{Ricky}. 
In order to minimize the occurrence of rebuffering events, HTTP streaming solutions such as Dynamic Adaptive Streaming over HTTP (DASH) allow their clients (or the video users) to adapt the video rate in accordance with the changing network conditions \cite{sodagar2011mpeg}. 
%provide an operative framework for media streaming over HTTP networks \cite{sodagar2011mpeg}. 
%DASH has become a popular choice of media streaming as most networks today are configured to operate based on HTTP/Transmission Control Protocol (TCP). 
Since the media delivery in DASH is based on reliable HTTP/TCP, there are no packet losses at the end user. %However, the network impairments such as congestion and packet drops can result in significant delays in the packet arrival at the end user. 
%Delays in the video packet arrival results in the emptying of the playback buffer 
%The playback is not resumed until sufficient content is available in the buffer. 
%The possibility of occurrence of rebuffering events in wireless networks is higher as the resources are scarce and are shared amongst multiple users. Due to resource sharing, the users can end up being starved for sufficient resources, thereby limiting the data rate and thus the throughput thereby increasing the chance for the occurrences of rebuffering events.
%The data rate of a wireless video user is highly influenced by the network dynamics such as the number of users, the load etc.
Rate adaptation is a key feature offered by the adaptive streaming frameworks that is useful in dynamic and time varying transmission environments such as mobile networks. 
However, the videos encoded at different rates offer different video qualities and therefore, rate adaptation results in a video quality that varies with time. Time varying video quality and rebuffering events can lead to significant degradation of the end user QoE \cite{LIVE_Netflix, LFOVIA_QoE}. 
%The QoE as perceived by the user is determined by a complex interplay of these distortions \cite{LFOVIA_QoE}. 
%As these QoE influencing events occur at different instants of time in a given video streaming session, the QoE becomes time varying in nature.
%QoE represents the degree of satisfaction of the users as they are subjected to time varying quality and rebuffering events.
Monitoring the continuous time QoE is vital for the optimal utilization of shared resources and thereby maximize the QoE of the video users in the network.
Continuous QoE evaluation is also useful in choosing the appropriate video rate so that the QoE degradations can be minimized. 
%Further, estimating the continuous QoE evolution can provide insights into the understanding of perceptual responses of the users when subjected to distortions, both individually as well as jointly.
%These form the motivating factors for our work.

In this paper, we make the following contributions:
\begin{itemize}
\item[1)] We propose an efficient method for measuring the continuous QoE of video streaming users based on a nonlinear state space (NLSS) model. The proposed model is based on the perceptual experience of the video streaming users unlike the network based QoE evaluation methods \cite{Ricky,Singh_QoE}.
\vspace{-0.01in}
\item[2)] We investigate three features for continuous QoE estimation, namely, (a) short time subjective quality, (b) playback indicator, and (c) time elapsed since the last rebuffering event \cite{NARX}. 
\item[3)] We conduct an evaluation of the proposed model on two continuous QoE databases and demonstrate a high QoE estimation performance of the proposed model outperforming the state-of-the-art QoE evaluation methods.
%To the best of our knowledge, this is the first work that conducts a comprehensive study of the proposed model over all publicly available continuous QoE databases for QoE estimation.
\end{itemize}

The rest of the paper is organized as follows. Section \ref{sec:background} gives a brief overview of the existing QoE modeling approaches. The proposed QoE model is presented in Section \ref{sec:nonlinear_state_space}. Section \ref{sec:QoE_estimation} describes the QoE evaluation methodology using the proposed approach. Performance evaluation and analysis of the proposed QoE model is discussed in Section \ref{sec:Results} followed by concluding remarks in Section \ref{sec:conclusions}.
%an overall QoE performance analysis in Section \ref{sec:overall_perf}. Finally, Section \ref{sec:conclusions} concludes the work.

\vspace{-0.15in}

\section{Related Work}
\label{sec:background}

\vspace{-0.05in}

QoE centric design has gained a lot of importance owing to several advantages to the multimedia service providers. 
Finding feature descriptors for prediction models that quantify the user QoE has been drawing a lot of attention lately \cite{IQX, Ricky, Tobias, Wang_ICIP2016}.
%Real time multimedia applications such as online video streaming require maintenance of an acceptable level of user QoE. 
%For providing a satisfactory quality of service, the end user QoE needs to be constantly monitored. 
Measuring the end user QoE is a challenging task as the QoE is highly subjective in nature. 
However, many subjective studies have shown that although individual preferences vary, by and large the QoE of users concurs to a particular trend \cite{Ricky,Tobias,TVSQ_Chen,LFOVIA_QoE,LIVE_Netflix}. 
%The subjective studies help a great deal in quantifying the QoE as they provide real time QoE data of the users in response to various QoE influencing events that are carefully designed for understanding the QoE behavioral patterns.

Video quality assessment (VQA) forms a crucial part of QoE estimation models in video streaming \cite{LFOVIA_QoE,NARX,TVSQ_Chen}. VQA has been studied in several works in the literature \cite{MSSSIM,MOVIE,STRRED,FLOSIM}. 
%The evolution of VQA metrics from traditional measures such as Peak Signal-to-Noise-Ratio is discussed in \cite{Winkler}. 
\cite{Chikkerur} provides a comprehensive study of various VQA metrics and suggests that the metrics MS-SSIM \cite{MSSSIM} and MOVIE \cite{MOVIE} provide good video quality prediction performances. 
An optical flow based VQA method proposed in \cite{FLOSIM} is shown to provide a superior video quality prediction performance over all the existing methods. Although VQA metrics incorporate the aspects that determine user's perceptual quality, they are insufficient for determining the QoE \cite{LFOVIA_QoE,TVSQ_Chen}. 
QoE is found to be determined not just by the video quality but also by a sequence of events occurring at different time instants in a video session such as rate adaptation and rebuffering. 
%Rate adaptation in adaptive streaming causes the video quality to fluctuate over time because of which the user QoE becomes time varying. 
%Nevertheless, VQA forms an integral part of most of the QoE estimation models \cite{LFOVIA_QoE}, \cite{NARX}, \cite{TVSQ_Chen}.

%In wireless networks, channel fluctuations cause the data rate to vary drastically due to mobility of the user, user dynamics, resource sharing etc. DASH allows the client to adapt the video rate to best match the data rate of the user. In spite of the best efforts, when the network/channel conditions degrade, the video client can run out of the playback content causing the playback to stall. Thus, the rate adaptation together with rebuffering events lead to a degradation of QoE.

There have been several efforts that address the challenge of QoE prediction for video streaming.
%\cite{Balachandran1}, \cite{Balachandran2}.
In \cite{Singh_QoE}, the authors identify some of the QoE metrics that are defined in the 3GPP DASH specification TS 26.247 standard. Some of them include the average throughput, initial playout delay, buffer level etc. However, these metrics can only act as indicators of the QoE and cannot measure the actual QoE as they do not capture the perceptual experience of the user.
There are other factors that have been identified as the factors affecting the QoE of a user such as the initial loading time, startup delay and so on \cite{Ricky,Tobias,LIVE_Mobile_Stall_II}.
However, it is shown in these studies that the startup delays have minimal or almost negligible impact on the QoE. 
%This suggests that the users are willing to wait for a considerable amount of time before the playback begins without causing any dissatisfaction. However, once the playback is begun, the QoE of a user is shown to become sensitive to time varying quality as well as rebuffering. 
Other QoE studies such as \cite{Ricky,LFOVIA_QoE,Tobias} indicate that the rebuffering events degrade the QoE severely. 
It is reported in these studies that the user is willing to sacrifice higher resolutions (or equivalently better quality) for avoiding interruptions in the playback. 
%It is reported in \cite{Pessemier} that the user is willing to sacrifice higher resolutions (or equivalently better quality) for avoiding interruptions in the playback. 
%Therefore, it becomes imperative for the video client to maintain sufficient content in the playback buffer at any point in time in order to avoid causing a severe QoE degradation for the users. 

In \cite{TVSQ_Chen}, Chen \textit{et al.} propose the Hammerstein Wiener model for measuring the perceptual time varying video quality due to rate adaptation. In \cite{Deepti_DQS}, Yeganeh \textit{et al.} propose the delivery quality score model to estimate the overall perceptual QoE due to rebuffering. 
It is to be noted that these methods study and model the time varying quality and rebuffering events separately and do not consider them jointly. In \cite{SQI}, Duanmu \textit{et al.} consider these QoE impairments jointly and design the streaming quality index to measure QoE. However, these QoE models evaluate only the overall QoE towards the end of watching a video and not the dynamic QoE of the users on a continuous time basis. There is a need for perceptually motivated continuous QoE estimation methods
%Such models are not suitable for QoE monitoring 
for the optimal utilization of network resources and thereby enhance the user QoE in real time. 
In \cite{LIVE_Netflix}, Bampis \textit{et al.} provide the LIVE Netflix Database along with a subjective study of user QoE in the presence of time varying quality and rebuffering together. 
Over this QoE database, a nonlinear autoregressive model (NARX) is proposed in \cite{NARX} based on an autoregressive neural network to estimate the continuous QoE.
In \cite{LFOVIA_QoE}, Eswara \textit{et al.} conduct a subjective study of continuous QoE on the videos at full high definition (FHD) and ultra high definition (UHD) resolutions and present the LFOVIA QoE Database. In addition, the authors also present a continuous QoE evaluation framework based on support vector regression (SVR-QoE). 
Although NARX and SVR-QoE modeling approaches address the continuous QoE estimation problem in the presence of both time varying quality and rebuffering, they are validated only on their respective QoE databases for which they are designed and proposed. We show in Section \ref{sec:Results} that each of the models' QoE estimation performance drops when trained and evaluated on other databases. Further, the QoE analysis of these models is not easily interpretable as they are built using machine learning algorithms.

Therefore, we propose a NLSS model for estimating continuous QoE that is more tractable for analysis. Using the proposed model, we conduct a comprehensive evaluation of the continuous QoE databases and show that the proposed model performs consistently well across the databases. We also show that the performance of the proposed QoE model is superior to the state-of-the-art QoE methods.
%Finally, we predict the overall QoE from the estimated continuous QoE scores using an average pooling strategy. It is found that a simple averaging is good enough to predict the overall QoE effectively.

\vspace{-0.1in}

\section{Nonlinear State Space Model}
\label{sec:nonlinear_state_space}

\vspace{-0.05in}

\begin{figure}
\centering
\includegraphics[scale=0.75]{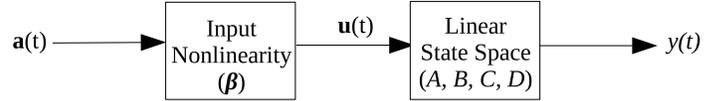}
\caption{Proposed nonlinear state space QoE model.}
\label{fig:nonlinear_SS}
\end{figure}

%The user QoE in video streaming is determined by the human visual perception to a large extent \cite{recency}. 
According to International Telecommunications Union, QoE is defined as the overall quality of an application or a service as \textit{perceived subjectively} by the end user \cite{ITU_QoE}. Many psycho-visual experiments conducted on the visual system suggest that the visual quality and the perceptual experience is highly nonlinear in nature due to nonlinear response properties of the neurons in the primary visual cortex \cite{MOVIE}. 
%Due to this, the QoE behavioral patterns of a user while watching a video involve nonlinearities. 
Further, it is observed through several subjective studies that the visual QoE 
%in general is dynamic and time-varying in nature, varying 
varies dynamically according to various QoE influencing events such as rate adaptation \cite{LIVE_Netflix, LFOVIA_QoE, TVSQ_Chen}. 
Such events result in the \textit{hysteresis effect} \cite{Kalpana_hysteresis}, where 
%, where in the past events leave a significant impact on the QoE at the current time instant. 
%This is particularly observed to be prominent in the cases where a poor visual quality occurring in the past, ripples through and leaves a significant impact on the current QoE, even though the rendered visual quality at the current instant is higher. 
%The hysteresis effect suggests that 
the continuous QoE involves a memory of a sequence of past events influencing the current QoE.
%In summary, the continuous QoE is nonlinear in nature and involves a memory of past events, influencing due to hysteresis effects. 
Thus, 
%in this section, we present a nonlinear state-space model to capture these effects and perform continuous QoE estimation. 
in the proposed NLSS model, the nonlinear properties of the neurons are captured using an explicit static nonlinear function and the memory effects are modeled using the state space design. 
%Since video streaming is a continuous process by nature, 
Fig. \ref{fig:nonlinear_SS} shows the proposed nonlinear dynamic QoE estimation model.
We evaluate the QoE on a continuous time basis using the proposed system. 

%The system model for QoE estimation includes an input nonlinearity followed by a linear state space process. 

Let $m$ be the number of inputs to the model. Let $\mathbb{R}^m_{\geq0}$ represent the set of all nonnegative real numbers in an $m$-dimensional space. Let $\textbf{a}(t) \in \mathbb{R}^m_{\geq0}$ represent the $m$-dimensional input feature vector to the system.
Let $\textbf{u}(t) \in \mathbb{R}^m_{\geq0}$ represent the time-indexed $m$-dimensional vector serving as the input to the linear state-space for estimating the QoE represented as $\hat{y}(t) \in \mathbb{R}$. 
Let \pmb{\textnormal{$\beta$}} = [$\beta_{11} \cdots \beta_{51}, \ \beta_{12} \cdots \beta_{52}, \ \cdots, \ \beta_{1m} \cdots \beta_{5m}$] represent the set of static nonlinear parameters of the model.
We define the nonlinearity as a sum of sigmoid function and linear function, as mentioned in the following.
%considered in Eq. (\ref{eq:INL})

\vspace{-0.1in}
\begin{equation}
u_i(t) = \frac{{\beta_3}_i}{1+exp(-({\beta_1}_ia_i(t)+{\beta_2}_i))} + {\beta_4}_ia_i(t) + {\beta_5}_i, \nonumber
\label{eq:INL}
\end{equation}
\begin{equation}
\forall i = 1, 2, \cdots, m. \nonumber
\end{equation}
%\boldsymbol{${\beta}$}
Let $\textbf{x}(t)$ be the state vector of the model at any time instant $t$. Using standard state space equations \cite{Ogata}, the output is given by
\begin{equation}
\hat{y}(t) = C\textbf{x}(t) + D\textbf{u}(t),
\label{eq:output}
\end{equation}
where, $C$ and $D$ are the output matrix and the feed-forward matrix, respectively. The state update equation is
\begin{equation}
\textbf{x}(t+1) = A\textbf{x}(t) + B\textbf{u}(t),
\label{eq:state_update}
\end{equation}
where, $A$ is the system matrix and $B$ is the input matrix. In our evaluation, the QoE estimation is performed every second, i.e., at a granularity of $t$ = 1 second. Next, we describe the states in the model.

\vspace{-0.1in}

\subsection{Identification of States}

%\vspace{-0.05in}

Let $\textbf{x} \in \mathbb{R}^s$, implying that the number of state variables is equal to $s$
%Let the number of state variables be $s$, implying that the dimension of the state vector $\textbf{x}(t)$ is equal to $s$. 
%In state-space models, 
and the state transitions are controlled by the input signals $\textbf{u}(t)$. Since there are $m$ such input signals, we set the number of state variables $s$ to be at least $m$ i.e., $s \geq m$.
%, so that hidden state transitions, if any, for improved estimation of QoE can be effectively captured.
Furthermore, we define the quantity $r$, $r > 0$ and $r \leq s$ such that 
%each of the $m$-dimensional inputs has a direct influence on $r$ states.
%In other words, at any instant of time $t$, a set of $r$ states are distinctly controlled by any one given input.
a set of $r$ states are controlled by each of the $m$-dimensional inputs distinctly.
Let these $r$ states be constituted by the previous $r$ values of each input.
Accordingly, the number of state variables $s$ is hence determined by the number of inputs which is equal to $m$ and the number of states $r$ corresponding to each input. Thus, we have the relation $s$ = $mr$. 
Here, $r$ represents the model order since it accounts for the previous inputs while making the state transition at any instant of time $t$. 
%By virtue of the construction of states, we have the following consequences in the state update equation. \\
%1) Rank($A$) = $s$ and Rank($B$) = $m$. \\
%2) Matrix $A \in \mathbb{R}^{s\times s}$ is block-diagonal, with sub-matrices constituting the diagonal $\in \mathbb{R}^{r\times r}$.
In addition, we impose the following constraints on the parameters of the state update equation: 1) Rank($A$) = $s$ and 
%2) $m$ $\leq$ Rank($B$) $\leq$ $s$. 
2) Rank($B$) = $m$.
These constraints are imposed in order to make the state space controllable \cite{Ogata}.

\vspace{-0.1in}

\subsection{Feature Selection}
\label{subsec:feat_selec}

%\vspace{-0.05in}

We consider three features as the input to the model as described in the following:

\begin{itemize}

\item[1)] \textit{Short Time Subjective Quality} (STSQ): STSQ of the current video segment can be calculated using any of the sophisticated video quality assessment (VQA) metrics. STSQ measures the perceptual video quality of the current video being rendered to the user.

\item[2)] \textit{Playback Indicator} (PI): 
%It is observed in the previous studies that the rebuffering results in a significant drop in the QoE \cite{LFOVIA_QoE}, \cite{LIVE_Netflix}, \cite{LIVE_Mobile_Stall_II}. Hence, 
Since rebufferings result in a significant drop in the QoE as indicated in several studies \cite{LIVE_Netflix,LFOVIA_QoE,LIVE_Mobile_Stall_II}, we employ a binary indicator variable PI to indicate whether the video currently is in the playback state or in the rebuffering state.
% Thus, PI can potentially serve as a feature to the state-space estimator.

\vspace{-0.05in}

\item[3)] \textit{Time elapsed since last rebuffering} ($\textnormal{T}_\textnormal{R}$): A rebuffering event is usually followed by a recovery phase, where the depreciated QoE due to rebuffering tries to recover as the playback progresses \cite{LFOVIA_QoE}. 
Therefore, we hypothesize that the improvement in QoE in the recovery phase is proportional to the time elapsed since the last rebuffering. 
Hence, we employ $\textnormal{T}_\textnormal{R}$ as an input feature to the model.
%Hence, we employ $\textnormal{T}_\textnormal{R}$ as an input feature.

\end{itemize}

\vspace{-0.05in}

\begin{figure}[t]
%\centering
\begin{subfigure}[b]{0.158\textwidth}
\includegraphics[scale=0.22]{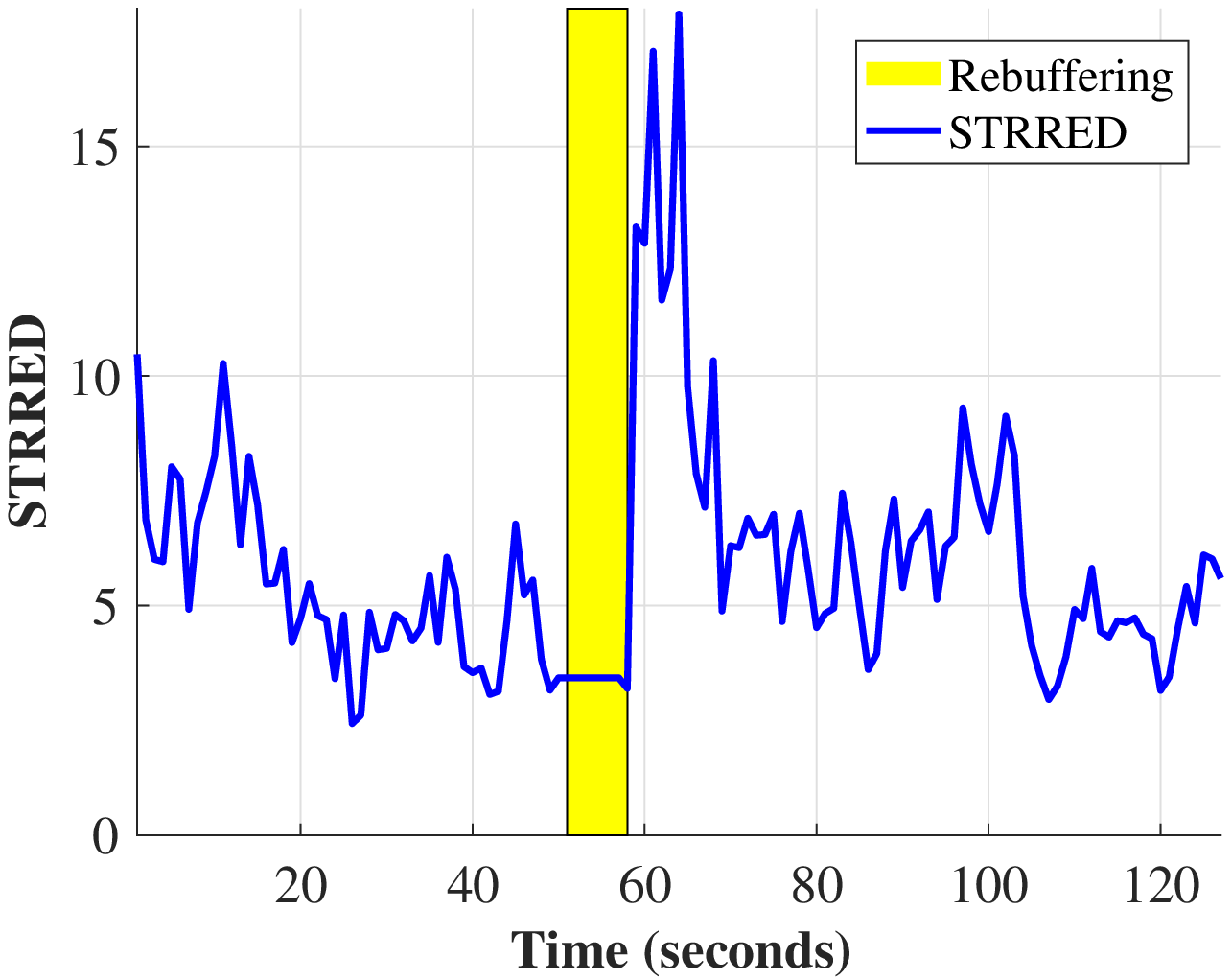}
\subcaption{\small STSQ}
\label{subfig:STSQ}
\end{subfigure}
%\hspace{0.40cm}
\begin{subfigure}[b]{0.158\textwidth}
\includegraphics[scale=0.22]{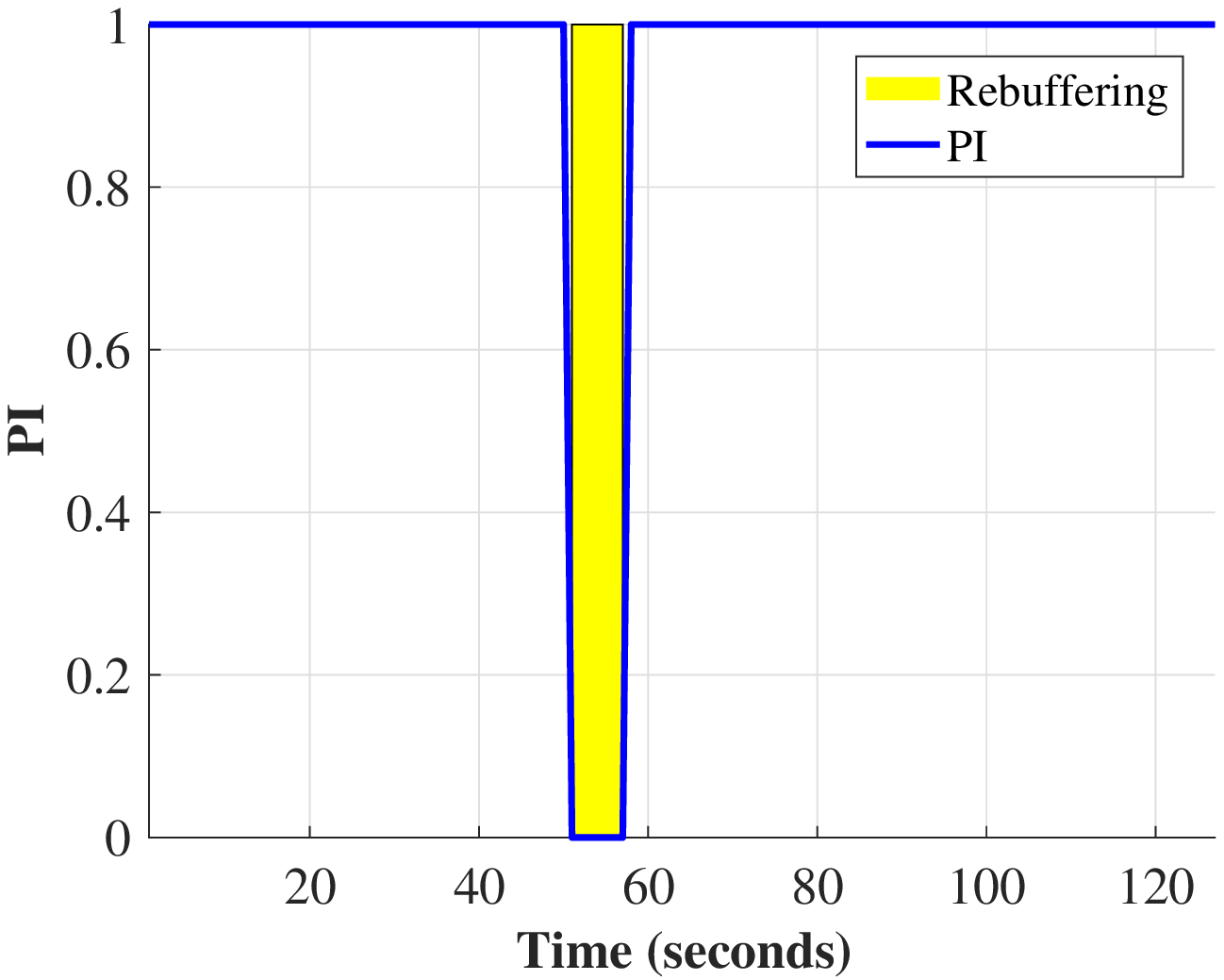}
\subcaption{\small PI}
\label{subfig:PI}
\end{subfigure}
%\centering
\begin{subfigure}[b]{0.158\textwidth}
\includegraphics[scale=0.22]{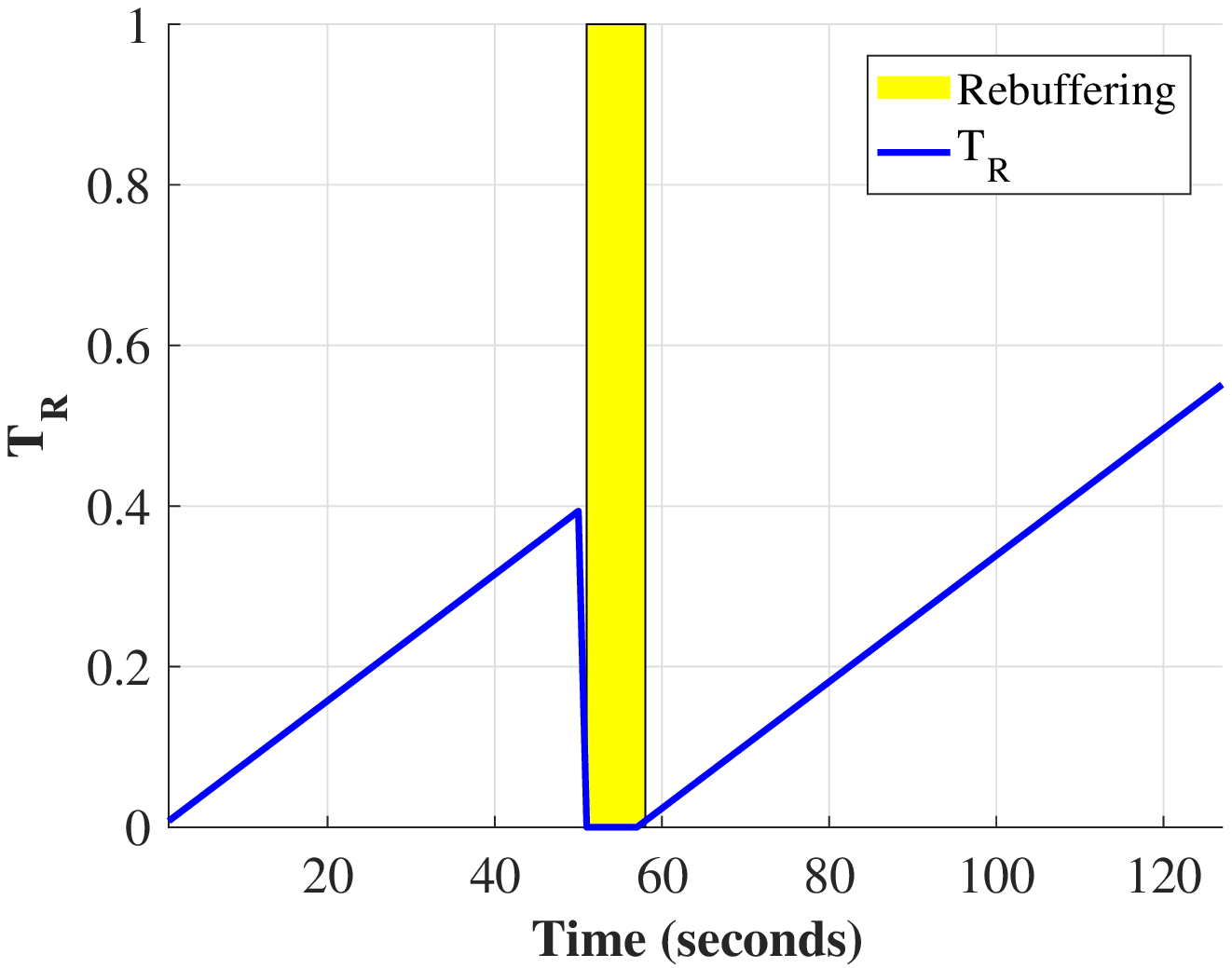}
\subcaption{\small $\textnormal{T}_\textnormal{R}$}
\label{subfig:TR}
\end{subfigure}
\caption{Illustration of the QoE features. Figs. \ref{subfig:STSQ}, \ref{subfig:PI}, and \ref{subfig:TR} depict the variation of features STSQ, PI, and $\textnormal{T}_\textnormal{R}$ with playback time, respectively, for an arbitrarily chosen video from the LFOVIA QoE Database \cite{LFOVIA_QoE}.}
%\medskip
%\small
%Figs. \ref{subfig:STSQ}, \ref{subfig:PI}, and \ref{subfig:TR} depict the variation of features STSQ, playback indicator, and $\textnormal{T}_\textnormal{R}$ with playback time, respectively, for an arbitrarily chosen video from the LFOVIA QoE Database \cite{LFOVIA_QoE}.
\label{fig:features_illustration}
\end{figure}

Fig. \ref{fig:features_illustration} illustrates the variation of the employed features STSQ, PI, and $\textnormal{T}_\textnormal{R}$ with playback time for one of the videos in the LFOVIA QoE Database \cite{LFOVIA_QoE}. Here, STRRED \cite{STRRED} is used as VQA for STSQ. STRRED shares an inverse relation with the video quality, i.e., a lower value of STRRED indicates a better video quality and vice-versa. 
In Fig. \ref{fig:features_illustration}, it can be observed that while STSQ tracks the time-variation in the quality, PI and $\textnormal{T}_\textnormal{R}$ are responsive to the rebuffering events.

Similar features have been employed in \cite{NARX} for QoE modeling as well. However, we would like to highlight that
only a limited set of features are available as part of the LIVE Netflix QoE Database \cite{LIVE_Netflix} upon which we evaluate the performance of our proposed model. Further, only a few videos of the database are made publicly available. This restricted us from the exploration and the investigation of furthermore QoE features, as the database is not available completely. 
Nevertheless, we demonstrate in Section \ref{sec:Results} that even with this set of limited features, the proposed QoE model is able to provide an excellent performance compared to the state-of-the-art QoE estimation methods.

\begin{figure*}[t]
\begin{subfigure}[b]{0.24\textwidth}
\centering
\includegraphics[scale=0.315]{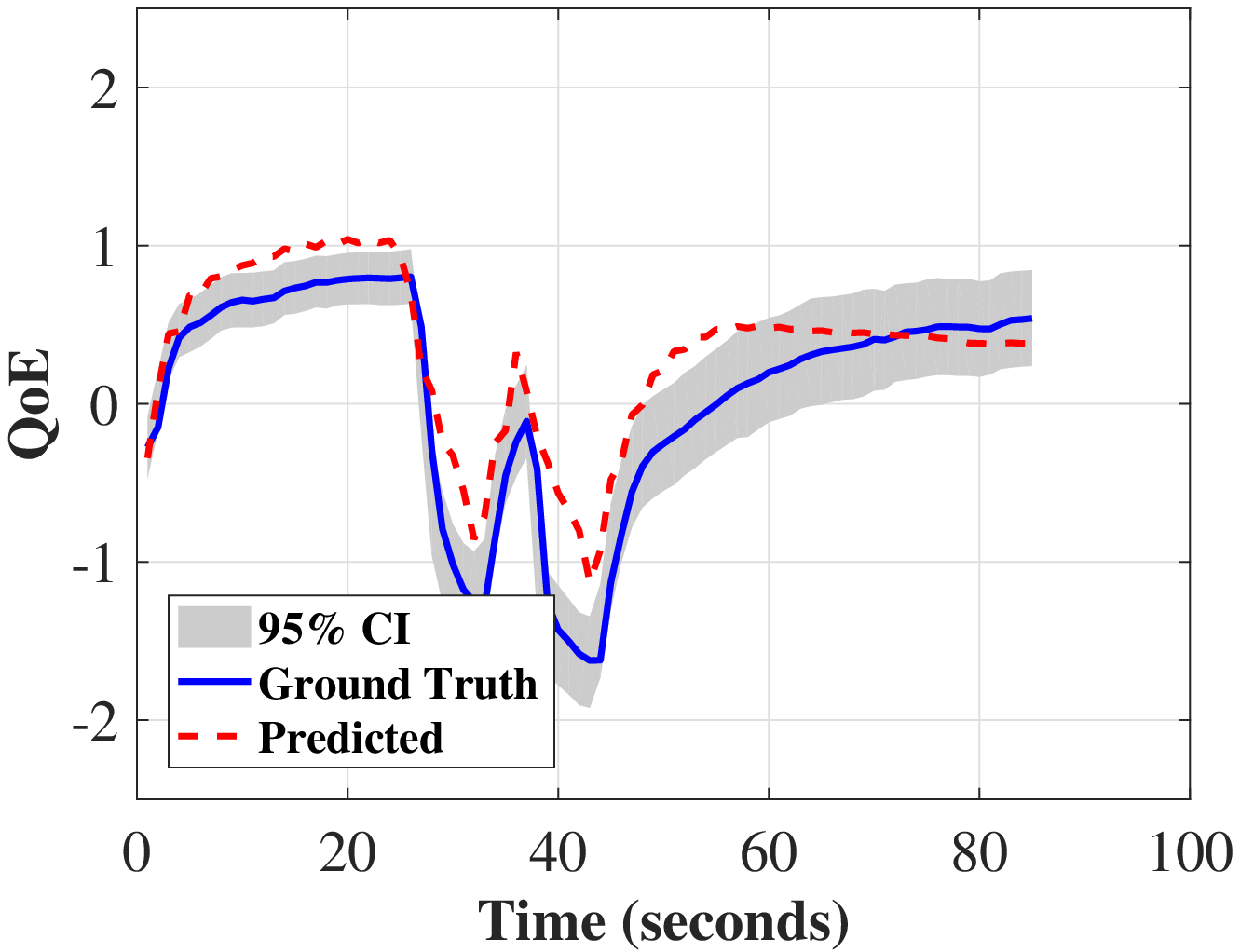}
\subcaption{\small Test Video 1}
\label{subfig:LIVE_Netflix_a}
\end{subfigure}
%\hspace{0.40cm}
\begin{subfigure}[b]{0.24\textwidth}
\centering
\includegraphics[scale=0.315]{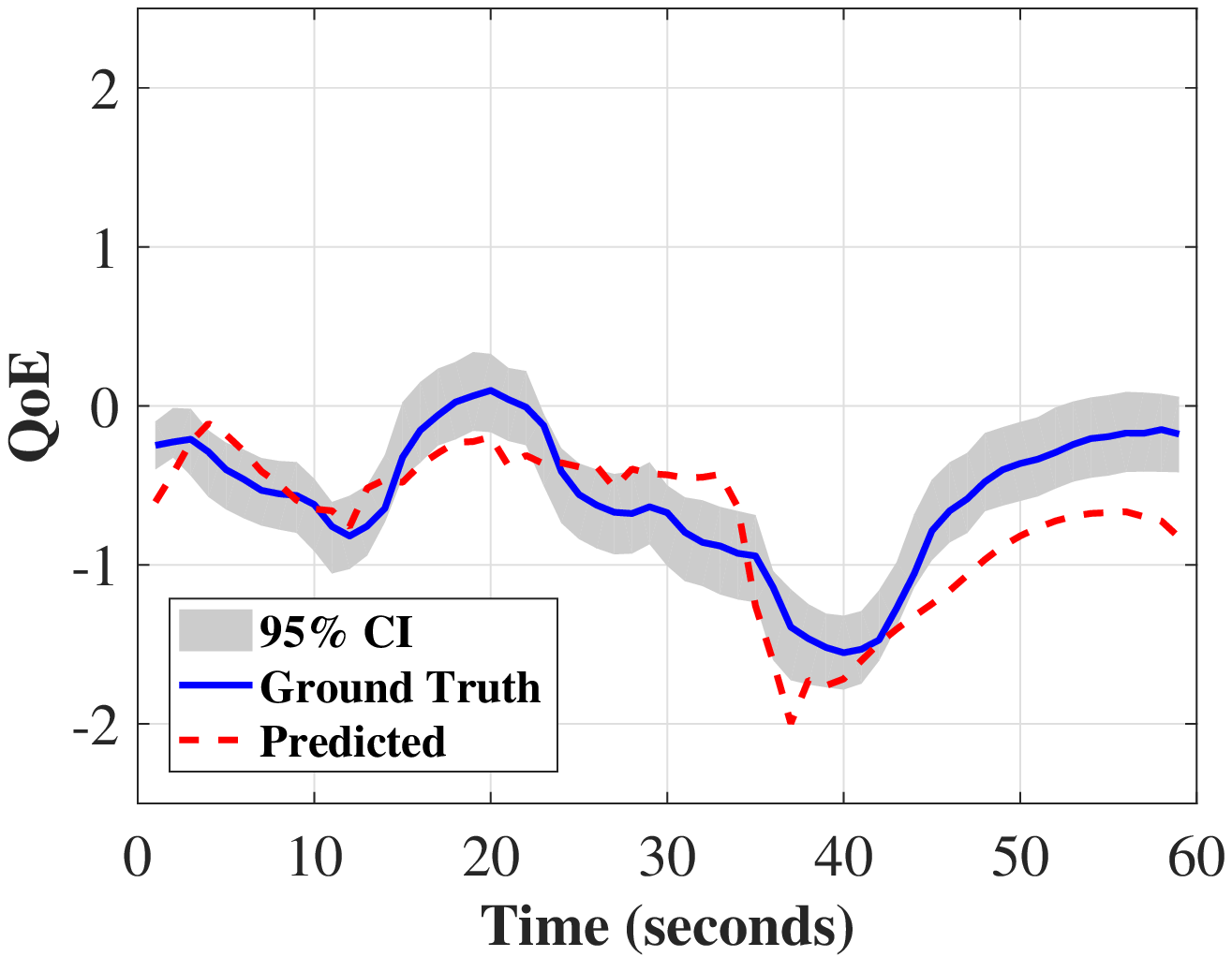}
\subcaption{\small Test Video 2}
\label{subfig:LIVE_Netflix_b}
\end{subfigure}
%\centering
\begin{subfigure}[b]{0.24\textwidth}
\centering
\includegraphics[scale=0.315]{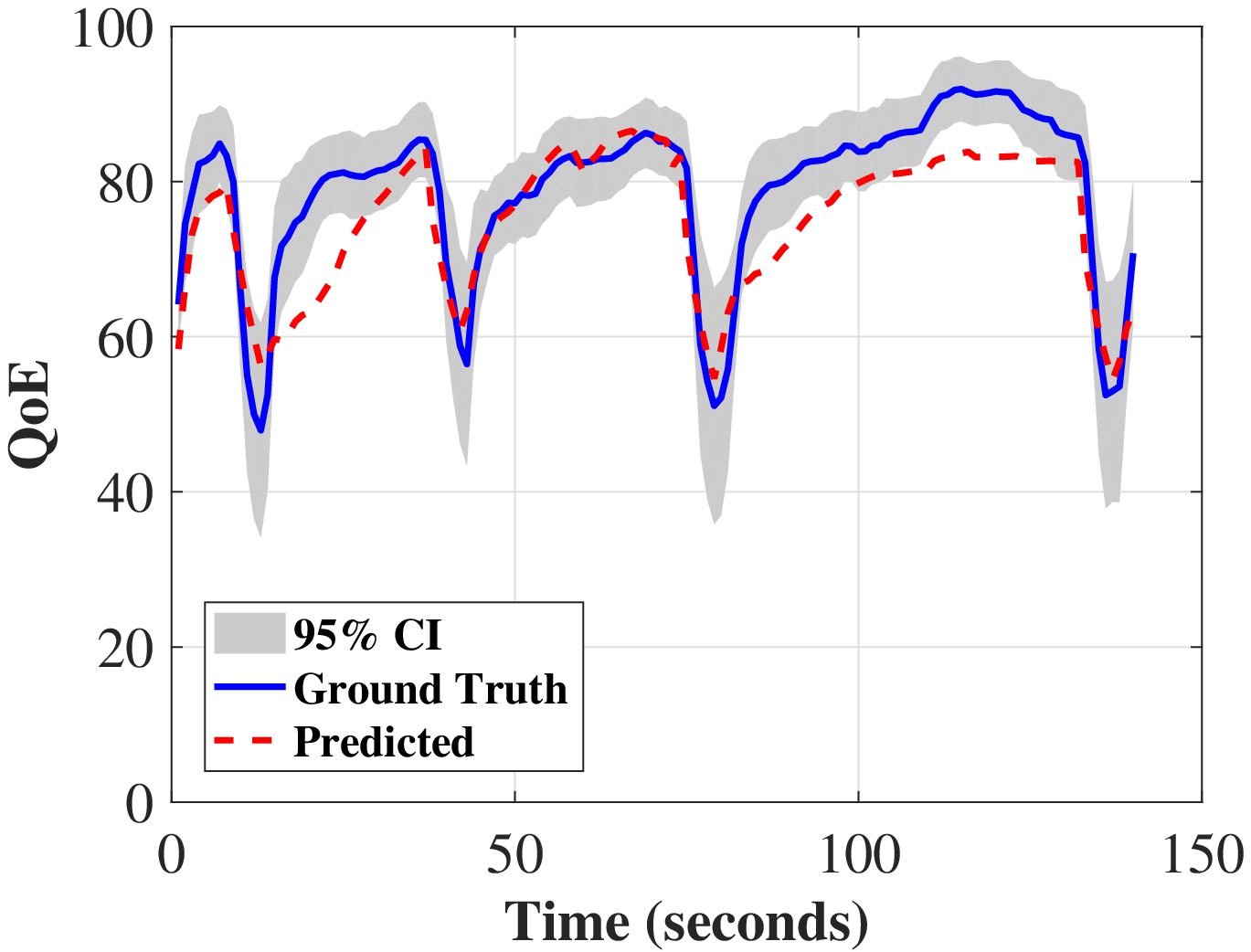}
\subcaption{\small Test Video 1}
\label{subfig:LFOVIA_QoE_a}
\end{subfigure}
\begin{subfigure}[b]{0.24\textwidth}
\centering
\includegraphics[scale=0.315]{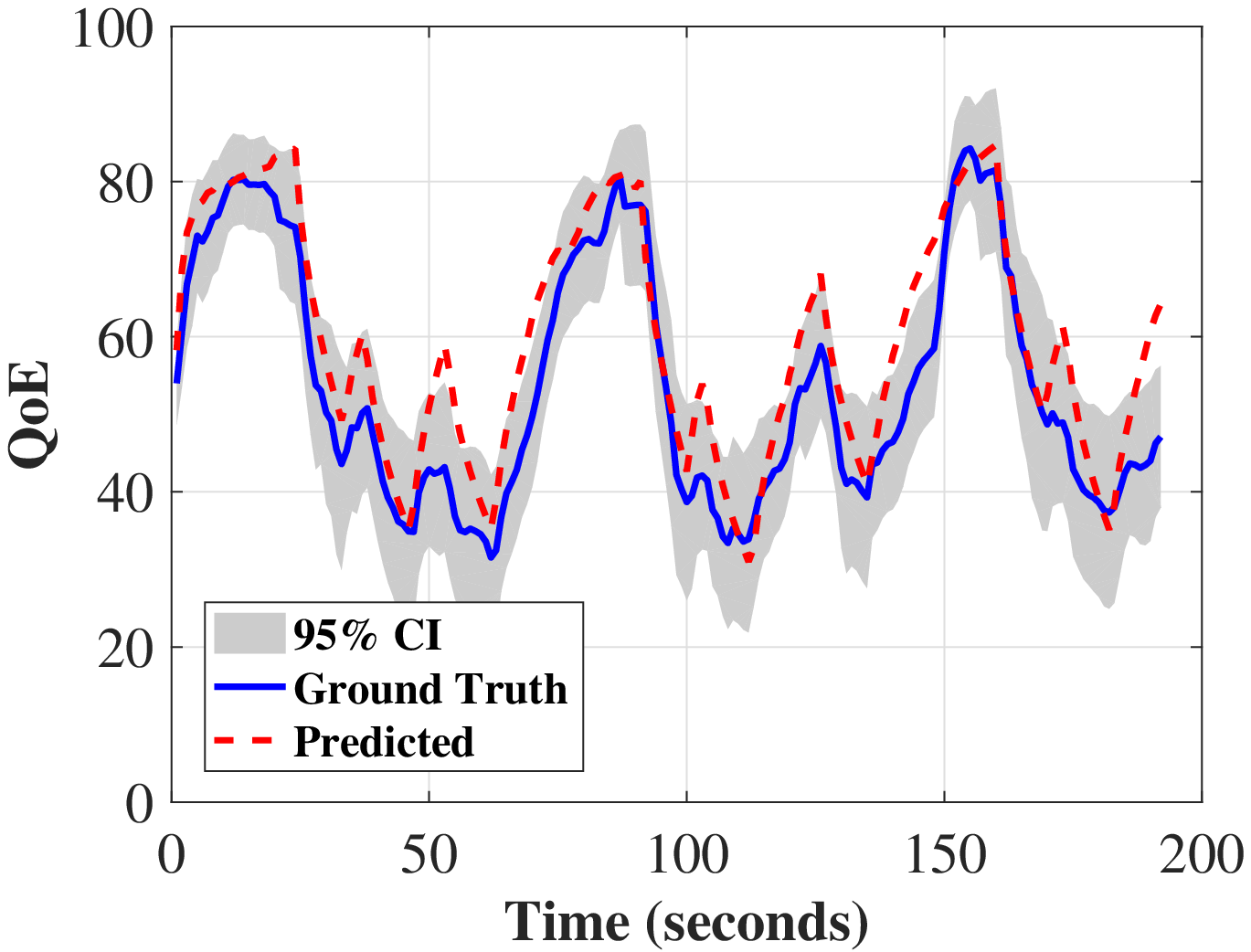}
\subcaption{\small Test Video 2}
\label{subfig:LFOVIA_QoE_b}
\end{subfigure}
\caption{QoE estimation performance of the proposed NLSS QoE model. Figs. \ref{subfig:LIVE_Netflix_a} and \ref{subfig:LIVE_Netflix_b} illustrate the performance of the proposed model over two abritrarily chosen test videos from the LIVE Netflix Database with STRRED as VQA metric for STSQ. Figs. \ref{subfig:LFOVIA_QoE_a} and \ref{subfig:LFOVIA_QoE_b} illustrate the performance of the proposed model over two abritrarily chosen test videos from the LFOVIA QoE Database with NIQE as VQA metric for STSQ.}
%using STRRED \cite{STRRED} as VQA for STSQ.}
%\medskip
%\small
%Figs. \ref{subfig:LIVE_Netflix_a} and \ref{subfig:LIVE_Netflix_b} illustrate the performance of the proposed model over two abritrarily chosen test videos from the LIVE Netflix Database with STRRED as VQA for STSQ.
%Figs. \ref{subfig:LFOVIA_QoE_a} and \ref{subfig:LFOVIA_QoE_b} illustrate the performance of the proposed model over two abritrarily chosen test videos from LFOVIA QoE Database with NIQE as VQA for STSQ.

\label{fig:performance_init}
\end{figure*}

\vspace{-0.1in}

\section{QoE Estimation}
\label{sec:QoE_estimation}

\vspace{-0.05in}

In this section, we describe the procedure for QoE evaluation using the proposed approach. Since there are three input features to the model, namely, STSQ, PI, and $\textnormal{T}_\textnormal{R}$, we have $m$ = 3. We consider the following VQA metrics for STSQ: 1) STRRED \cite{STRRED}, 2) MS-SSIM \cite{MSSSIM}, 3) PSNR \cite{Kalpana}, and (4) NIQE \cite{NIQE}. PI and $\textnormal{T}_\textnormal{R}$ are simple dynamic features that can be obtained directly by tracking the status of the playback.
Further, it has been observed in the previous studies that the user QoE is heavily influenced by the past experience of about 2-3 seconds \cite{TVSQ_Chen,C3D_TVSQ}. Hence, we set the model order $r$ = 3 implying the state space dimension $s$ = $mr$ = 9. 
%The $s$-dimensional vector $\textbf{x}(t)$ is a constantly evolving state vector and holds the dynamic state of the QoE estimator at any instant in time.

The proposed model is trained using the videos from the training set and evaluated for its performance on the test set. We consider non-overlapping training and test sets in all our evaluations.
%The QoE model is developed based on training and test procedure, where in each database is divided into a non-overlapping training set and the test set. 
During training, the nonlinearity function parameters \pmb{\textnormal{$\beta$}} and the state space parameters, namely the matrices $A$, $B$, $C$ and $D$ are determined by performing least squares minimization between the ground truth QoE and the estimated QoE.
% using Levenberg-Marquardt algorithm.
%The trained model is then tested on the videos from the test set. 
While evaluating the trained model on the test set, it must be noted that there are two unknowns to be determined as per (\ref{eq:output}) - the state vector $\textbf{x}$ and the output QoE $\hat{y}$. Since the interest of evaluation is the QoE $\hat{y}$, the state vector $\textbf{x}$ must be initialized with an appropriate initial state \textbf{x}(0). 
To overcome this problem, we resort to training data based state initialization methodology where in
%discussed in the following subsection.
the state of the model is initialized
based on the feature-QoE pair of the videos that are used in the training process. 
The best state initializer 
%for the input-output pair 
for the training set is determined and is subsequently used for the evaluation of the test video.

\vspace{-0.1in}

\section{Performance Evaluation and Analysis}
\label{sec:Results}

\vspace{-0.05in}

In this section, we discuss the performance evaluation of the proposed model on the QoE databases. We also evaluate the performance of the linear state space (LSS) by excluding the nonlinearity in Fig. \ref{fig:nonlinear_SS}. Both LSS and NLSS models are compared against the state-of-the-art QoE methods.

The performance of QoE estimation using the proposed model is quantified using the following three measures: 1) Linear Correlation Coefficient (LCC), 2) Spearman Rank Order Correlation Coefficient (SROCC), and 3) Normalized Root Mean Squared Error ($\textnormal{RMSE}_\textnormal{n}$).
Since the QoE databases have different QoE score ranges, we normalize the actual RMSE values to obtain `$\textnormal{RMSE}_\textnormal{n}$'. 
For a good performing model, LCC and SROCC values should be higher and $\textnormal{RMSE}_\textnormal{n}$ should be as low as possible.

We investigate the performance of the proposed model over two publicly available continuous QoE databases: 1) LIVE Netflix Database \cite{LIVE_Netflix} and 2) LFOVIA QoE Database \cite{LFOVIA_QoE}. 
%In our performance evaluation, we ensure that the training and test data do not overlap.

\begin{table}[t]
\caption{Performance of the proposed QoE model over the LIVE Netflix Database \cite{LIVE_Netflix} under various VQA metrics for STSQ. Text in italics indicates the state-of-the-art QoE model. The best performing results are shown in bold.}
\centering
\begin{tabular}{c|c|c|c|c} \hline
\label{table:LIVE_Netflix_init}

\textbf{QoE Model} & \textbf{VQA} & \textbf{LCC} & \textbf{SROCC} & \textbf{RMSE}$_\textnormal{n}$\textbf{(}\%\textbf{)}  \\ \hline

\multirow{4}{*}{NLSS} & STRRED \cite{STRRED} & \textbf{0.66} & 0.48 & 16.09 \\ \hhline{~----}

& MS-SSIM \cite{MSSSIM} & 0.58 & 0.42 & 18.22  \\ \hhline{~----}

& PSNR \cite{Kalpana} & 0.47 & 0.33 & 24.29 \\ \hhline{~----}

& NIQE \cite{NIQE} & 0.53 & 0.30 & 14.50  \\ \hline

LSS & STRRED \cite{STRRED} & 0.57 & 0.44 & 19.20 \\ \hline

\textit{NARX} \cite{NARX} & \textit{STRRED} \cite{STRRED} & 0.62 & \textbf{0.56} & \textbf{8.52} \\ \hline

\end{tabular}
\end{table}

\begin{table}[t]
\caption{Performance of the proposed QoE model over the LFOVIA QoE Database \cite{LFOVIA_QoE} under various VQA metrics for STSQ. 
%using the state initializer derived from the training. 
Text in italics indicates the state-of-the-art QoE models. The best performing results are shown in bold.}
\centering
\begin{tabular}{c|c|c|c|c} \hline
\label{table:LFOVIA_QoE_init}

\textbf{QoE Model} & \textbf{VQA} & \textbf{LCC} & \textbf{SROCC} & \textbf{RMSE}$_\textnormal{n}$\textbf{(}\%\textbf{)}  \\ \hline

\multirow{4}{*}{NLSS} & STRRED \cite{STRRED} & 0.77 & 0.69 & 7.59 \\ \hhline{~----}

& MS-SSIM \cite{MSSSIM} & 0.78 & 0.68 & 7.37  \\ \hhline{~----}

& PSNR \cite{Kalpana} & 0.02 & 0.08 & 8119  \\ \hhline{~----}

& NIQE \cite{NIQE} & \textbf{0.83} & \textbf{0.79} & \textbf{6.97}  \\ \hline

%\textit{NARX} \cite{NARX} & \multirow{2}{*}{0.67} & \multirow{2}{*}{0.60} & \multirow{2}{*}{9.34}  \\
%(\textit{STRRED}) & & &  \\ \hline

LSS & NIQE \cite{NIQE} & 0.78 & 0.69 & 7.53  \\ \hline

\textit{NARX} \cite{NARX} & \textit{NIQE} \cite{NIQE} & 0.75 & 0.69 & 7.87  \\ \hline

%\textit{SVR-QoE} \cite{LFOVIA_QoE} & \multirow{2}{*}{0.69} & \multirow{2}{*}{0.65} & \multirow{2}{*}{10.44} \\ 
%(\textit{STRRED}) &  &  & \\ \hline

\textit{SVR-QoE} \cite{LFOVIA_QoE} & \textit{NIQE} \cite{NIQE} & 0.79 & 0.75 & 8.32 \\ \hline

\end{tabular}
\end{table}

\vspace{-0.1in}

\subsection{LIVE Netflix Database}

\vspace{-0.05in}

We employ a standardized training and testing procedure with a training-test split as described in \cite{NARX}.
Accordingly, only one video in the database is considered in the test set in each training-test split. The model is trained using the videos that do not have the same content and the playout pattern as of the video in the test set. This procedure is repeated for all the videos in the database as the test set.
Table \ref{table:LIVE_Netflix_init} presents the QoE estimation performance of the proposed model.
Figs. \ref{subfig:LIVE_Netflix_a} and \ref{subfig:LIVE_Netflix_b} show the QoE estimation performance on sample test set videos of the database.

\vspace{-0.1in}

\subsection{LFOVIA QoE Database}

\vspace{-0.05in}

A training-test procedure similar to that of the LIVE Netflix Database is employed for QoE evaluation on the LFOVIA QoE Database, where the videos having the playout pattern same as that of the test video are excluded from training.
Table \ref{table:LFOVIA_QoE_init} presents the QoE estimation performance of the proposed model 
%using the state initializer obtained from training 
over the LFOVIA QoE Database. 
Figs. \ref{subfig:LFOVIA_QoE_a} and \ref{subfig:LFOVIA_QoE_b} show the QoE estimation performance on sample test set videos of the database.

%It can be observed that the QoE estimation performance is inferior to the case discussed previously. Although the model parameters are same in both cases, the initial states are essentially different. 
%A difference in the performance indicates that the state space initialization is important for the QoE estimation.
From Figs. \ref{subfig:LIVE_Netflix_a} and \ref{subfig:LIVE_Netflix_b}, it can be observed that though there is a gap between the estimated and the ground truth QoE, the trend in the QoE evolution appears to be similar and coherent. This is reflected in terms of higher LCC performance reported in Table \ref{table:LIVE_Netflix_init}, although the proposed approach yields a performance slightly inferior to NARX \cite{NARX} in terms of SROCC and RMSE$_\textnormal{n}$ on the LIVE Netflix Database. However, it is to be noted that the performance of the proposed model is achieved using a model order of 3, unlike the NARX approach which requires higher model orders (of the order of 15) to achieve a similar performance. This reduction in the model order significantly lowers the computational complexity of the QoE estimator.
%This can be noted from the Table \ref{table:LIVE_Netflix_init} as well. Even though the ORs are larger, the correlation between the two time series is higher. A better state initializer would reduce this gap and can make the estimation better further.

From Figs. \ref{subfig:LFOVIA_QoE_a} and \ref{subfig:LFOVIA_QoE_b}, and Table \ref{table:LFOVIA_QoE_init}, it can be observed that the QoE estimation performance using the proposed approach is superior when compared to the state-of-the-art methods NARX \cite{NARX} and SVR-QoE \cite{LFOVIA_QoE} in terms of all the performance measures on the LFOVIA QoE Database. Although NARX performs well on the LIVE Netflix Database,
%on which it is designed and proposed,
its QoE prediction performance is inferior on the LFOVIA QoE Database. On the other hand, the proposed QoE model provides a comparable performance on the LIVE Netflix Database and a superior performance on the LFOVIA QoE database. These results demonstrate the efficacy of the employed features for QoE estimation, thereby substantiating the hypothesis presented in Section \ref{subsec:feat_selec}.
%despite employing the state space initializer obtained from the training videos. 

It can be observed that STRRED emerges as the best performing VQA metric for STSQ on the LIVE Netflix Database whereas NIQE is found to be the best performing VQA metric for STSQ on the LFOVIA QoE Database. PSNR performs the least of all VQAs as it is not a perceptual VQA/QoE metric.
It can be noted that different VQA metrics yield varying QoE performances across the two databases.
This could be attributed to the
%the performance capability of STSQ, which in turn depends on the VQA method used for computing STSQs.
%The difference in the performance could be attributed to 
VQA metrics' ability to predict the video quality at different resolutions. Video resolution is an important aspect while measuring the video quality. All of the considered metrics for STSQ are demonstrated to perform well at resolutions lower than high definition. However, their VQA performance on videos having higher resolutions such as FHD and UHD (which is actually the case in the considered databases) is unknown \cite{LFOVIA_QoE}. This suggests the need for sophisticated VQA metrics that can provide excellent quality prediction performance consistently across all resolutions.
Such a VQA metric can be directly employed as the proposed model provides enough flexibility 
%allowing the QoE estimation systems to 
to incorporate appropriate VQA metric of choice for QoE evaluation.
%Thus, we conclude that the proposed nonlinear state space model is an effective approach for continuous QoE estimation.

A comparison between LSS and NLSS approaches shows a clear improvement in the QoE estimation performance with the addition of the nonlinearity. However, it is interesting to note that even a LSS system is able to achieve a performance comparable to that of the state-of-the-art QoE methods. Therefore, the LSS QoE model can serve as baseline for comparison with nonlinear QoE modeling approaches.  Thus, the proposed approach using state space provides a new and promising perspective for continuous QoE modeling and design of QoE centric networks.
%enabling analysis in real time.

\vspace{-0.1in}

\subsection{Controllability and Observability Analysis}

\vspace{-0.05in}

Since the state transitions in the model are driven by the input signal $\textbf{u}$, it is important to check for the state dynamics in order to ensure that the states do not enter into an undesired state due to spurious transitions or end up being in a deadlock. 
%Therefore, in this subsection, we discuss the aspects of controllability and observability of the LSS in the proposed model. 
Hence, we investigate the controllability of the LSS by examining the rank of the controllability matrix
$[B | AB | \cdots | A^{s-1}B]$ as in \cite{Ogata}. In our analysis of the trained models, it is found that the controllability matrix is full rank with rank being equal to $s$ in all cases of training implying that the system is completely state controllable. Similarly, the rank of the observability matrix $[C^T | A^TC^T | \cdots | (A^T)^{s-1}C^T]$ is also found to be full rank in all training cases, implying that the system is completely observable \cite{Ogata}.

\vspace{-0.1in}

\section{Conclusions and Future Work}
\label{sec:conclusions}

\vspace{-0.05in}

In this paper, we presented a nonlinear state space model for continuous video QoE evaluation. The proposed model predicts the QoE continuously as the user watches videos that involve time-varying qualities and interruptions in the playback due to rebuffering, that is typical of an HTTP streaming scenario. We studied the QoE behavioral patterns from two publicly available continuous QoE databases. We modeled the evolution of user QoE using state transitions that are triggered by a set of QoE influencing dynamic input features. The proposed QoE model was trained and evaluated on these two databases. On LFOVIA QoE Database, the proposed model outperformed all state-of-the-art QoE models. On LIVE Netflix Database, the proposed model showed a competitive performance and outperformed the state-of-the-art QoE model for the LCC performance measure. It must be noted here that this performance was achieved using a model order of 3, unlike the NARX approach which requires higher model orders to achieve a similar performance, thus significantly reducing the computational complexity of the QoE evaluation system. A reasonable performance of the proposed LSS QoE model suggests that the simplified model can be used as baseline for evaluating the performance of nonlinear QoE modeling approaches.
% over LIVE Netflix Database, it outperforms  demonstrating the effectiveness of the proposed approach. 
%In addition, the QoE prediction performance obtained using the proposed approach is observed to be quite consistent across databases. 
%Further, the state space initialization analysis illustrates that using the proposed model, a carefully initialized state can lead to an extremely good QoE prediction. 
The proposed QoE model is verified for both controllability and observability, validating the robustness of the model. In future, we intend to extend this model to investigate the stochastic properties of the state space for QoE analysis.
%Finally, an overall QoE prediction performance analysis suggests that the mean and the median continuous QoE pooling strategies is effective in quantifying the overall satisfactory levels of the users.
% in case of sessions that involve both time-varying video qualities and rebufering events.

%We presented a comprehensive model for QoE estimation based on state-space model. None of the previous works have considered or demonstrated the performance of their QoE estimation models over all the available databases. To the best of our knowledge, we consider all the databases to demonstrate the problem of continuous QoE estimation.
%It is observed that the evolution of QoE dynamics is linear in nature. However, the QoE estimation itself is not linear. The nonlinearities involved in the perceptual QoE is captured in STSQ.

%\section*{Acknowledgment}
%
%The authors would like to thank Dr. Nandini Ramesh Sankar of IIT Hyderabad for the suggestions that helped in significantly improving the quality and presentation of the paper.

\vspace{-0.1in}

\bibliographystyle{IEEEbib}
\bibliography{SS_QoE_arxiv}

\begin{thebibliography}{10}

\bibitem{cisco}
Cisco,
\newblock ``Cisco visual networking index: Global mobile data traffic forecast
  update, 2016--2021,''
\newblock {\em Cisco White Paper 1454457600805266}, Feb. 2017.

\bibitem{Ricky}
R.~K. P.~Mok \textit{et al}.,
\newblock ``Measuring the quality of experience of http video streaming,''
\newblock in {\em Proc. IFIP/IEEE IM}, May 2011, pp. 485--492.

\bibitem{sodagar2011mpeg}
Iraj Sodagar,
\newblock ``The mpeg-dash standard for multimedia streaming over the
  internet,''
\newblock {\em IEEE MultiMedia}, vol. 18, no. 4, pp. 62--67, Apr. 2011.

\bibitem{LIVE_Netflix}
C.~G.~Bampis \textit{et al}.,
\newblock ``Study of temporal effects on subjective video quality of
  experience,''
\newblock {\em IEEE Transactions on Image Processing}, vol. 26, no. 11, pp.
  5217--5231, Nov. 2017.

\bibitem{LFOVIA_QoE}
N.~Eswara \textit{et al.},
\newblock ``A continuous qoe evaluation framework for video streaming over
  http,''
\newblock {\em IEEE Trans. Circuits Syst. Video Technol.}, vol. PP, no. 99, pp.
  1--1, 2017.

\bibitem{Singh_QoE}
O.~Oyman and S.~Singh,
\newblock ``Quality of experience for http adaptive streaming services,''
\newblock {\em IEEE Communications Magazine}, vol. 50, no. 4, pp. 20--27, Apr.
  2012.

\bibitem{NARX}
C.~G.~Bampis \textit{et al}.,
\newblock ``Continuous prediction of streaming video qoe using dynamic
  networks,''
\newblock {\em IEEE Signal Process. Lett.}, vol. 24, no. 7, pp. 1083--1087,
  Jul. 2017.

\bibitem{IQX}
M.~Fiedler \textit{et al}.,
\newblock ``A generic quantitative relationship between quality of experience
  and quality of service,''
\newblock {\em IEEE Netw.}, vol. 24, no. 2, pp. 36--41, Mar. 2010.

\bibitem{Tobias}
Tobias~Ho{\ss}feld \textit{et al}.,
\newblock {\em Data Traffic Monitoring and Analysis: From Measurement,
  Classification, and Anomaly Detection to Quality of Experience}, chapter~13,
  pp. 264--301,
\newblock Springer Berlin Heidelberg, 2013.

\bibitem{Wang_ICIP2016}
K.~Zeng \textit{et al}.,
\newblock ``Quality-of-experience of streaming video: Interactions between
  presentation quality and playback stalling,''
\newblock in {\em Proc. IEEE ICIP}, Sep. 2016, pp. 2405--2409.

\bibitem{TVSQ_Chen}
Chao~Chen \textit{et al}.,
\newblock ``Modeling the time-varying subjective quality of http video streams
  with rate adaptations,''
\newblock {\em IEEE Trans. Image Process.}, vol. 23, no. 5, pp. 2206--2221, May
  2014.

\bibitem{MSSSIM}
Z.~Wang \textit{et al}.,
\newblock ``Multiscale structural similarity for image quality assessment,''
\newblock in {\em Proc. Asilomar Conf. on Signals, Systems and Computers}, Nov.
  2003, vol.~2, pp. 1398--1402.

\bibitem{MOVIE}
K.~Seshadrinathan and A.~C. Bovik,
\newblock ``Motion tuned spatio-temporal quality assessment of natural
  videos,''
\newblock {\em IEEE Trans. Image Process.}, vol. 19, no. 2, pp. 335--350, Feb.
  2010.

\bibitem{STRRED}
Rajiv Soundararajan and Alan~C Bovik,
\newblock ``Video quality assessment by reduced reference spatio-temporal
  entropic differencing,''
\newblock {\em IEEE Trans. Circuits Syst. Video Technol.}, vol. 23, no. 4, pp.
  684--694, Apr. 2013.

\bibitem{FLOSIM}
M.~K. and S.~S. Channappayya,
\newblock ``An optical flow-based full reference video quality assessment
  algorithm,''
\newblock {\em IEEE Trans. Image Process.}, vol. 25, no. 6, pp. 2480--2492,
  Jun. 2016.

\bibitem{Chikkerur}
S.~Chikkerur \textit{et al}.,
\newblock ``Objective video quality assessment methods: A classification,
  review, and performance comparison,''
\newblock {\em IEEE Trans. Broadcast.}, vol. 57, no. 2, pp. 165--182, Jun.
  2011.

\bibitem{LIVE_Mobile_Stall_II}
D.~Ghadiyaram \textit{et al}.,
\newblock ``A subjective and objective study of stalling events in mobile
  streaming videos,''
\newblock {\em IEEE Trans. Circuits Syst. Video Technol.}, vol. PP, no. 99, pp.
  1--1, 2017.

\bibitem{Deepti_DQS}
H.~Yeganeh \textit{et al}.,
\newblock ``Delivery quality score model for internet video,''
\newblock in {\em Proc. IEEE ICIP}, Oct. 2014, pp. 2007--2011.

\bibitem{SQI}
Z.~Duanmu \textit{et al}.,
\newblock ``A quality-of-experience index for streaming video,''
\newblock {\em IEEE J. Sel. Topics Signal Process.}, vol. 11, no. 1, pp.
  154--166, Feb. 2017.

\bibitem{ITU_QoE}
ITU,
\newblock ``Quality of experience requirements for iptv services,''
\newblock {\em Recommendation ITU-T G.1080}, Dec. 2008.

\bibitem{Kalpana_hysteresis}
K.~Seshadrinathan and A.~C. Bovik,
\newblock ``Temporal hysteresis model of time varying subjective video
  quality,''
\newblock in {\em Proc. IEEE ICASSP}, May 2011, pp. 1153--1156.

\bibitem{Ogata}
Katsuhiko Ogata,
\newblock {\em Modern Control Engineering},
\newblock Prentice Hall PTR, NJ, USA, 5th edition, 2010.

\bibitem{Kalpana}
K.~Seshadrinathan \textit{et al}.,
\newblock ``Study of subjective and objective quality assessment of video,''
\newblock {\em IEEE Trans. Image Process.}, vol. 19, no. 6, pp. 1427--1441,
  Jun. 2010.

\bibitem{NIQE}
A.~Mittal \textit{et al}.,
\newblock ``Making a ``completely blind" image quality analyzer,''
\newblock {\em IEEE Signal Process. Lett.}, vol. 20, no. 3, pp. 209--212, Mar.
  2013.

\bibitem{C3D_TVSQ}
N.~Eswara \textit{et al}.,
\newblock ``A linear regression framework for assessing time-varying subjective
  quality in http streaming,''
\newblock in {\em Proc. IEEE GlobalSIP}, Nov 2017, pp. 31--35.

\end{thebibliography}

\end{document}